# Covalent features in the hydrogen bond of a water dimer: molecular orbital analysis

Bo Wang[1,2], Wanrun Jiang[1,2], Xing Dai[1,2], Yang Gao[1,2], Zhigang Wang[1,2,3,*] and Rui-Qin Zhang[4,5,*]

[1]Institute of Atomic and Molecular Physics, Jilin University, Changchun 130012, China

[2]Jilin Provincial Key Laboratory of Applied Atomic and Molecular Spectroscopy (Jilin University), Changchun 130012, China

[3]Institute of Theoretical Chemistry, Jilin University, Changchun 130023, China

[4]Department of Physics and Materials Science and Centre for Functional Photonics (CFP), City University of Hong Kong, Hong Kong SAR, China

[5]Beijing Computational Science Research Center, Beijing 100084, China.

*To whom correspondence should be addressed. E-mail: wangzg@jlu.edu.cn, Tel.: +86-431-85168817; aprqz@cityu.edu.hk, Tel.: +852-34427849.

**Abstract**: The covalent-like characteristics of hydrogen bonds offer a new perspective on intermolecular interactions. Here, using density functional theory and post-Hartree-Fock methods, we reveal that there are two bonding molecular orbitals (MOs) crossing the hydrogen-bond's O and H atoms in the water dimer. Energy decomposition analysis also shows a non-negligible contribution of the induction term. These results illustrate the covalent-like character of the hydrogen bond between water molecules, which contributes to the essential understanding of ice, liquid water, related materials, and life sciences.

**Introduction**

Hydrogen bonding (H-bonding) is an essential interaction in nature and plays a crucial role in materials and life sciences. The water dimer $(H_2O)_2$ is one of the most typical models for studying the H-bonding system and, as such, much scientific effort has been directed toward understanding its properties.[1-3] Several studies have indicated that H-bonds have covalent-like characteristics.[4-6] Recent experiments have not only revealed that the covalent-like characteristics of H-bonds exist between two 8-hydroxyquiline molecules assembled on a Cu(111) substrate,[7] but also directly visualized the frontier MOs of adsorbed water.[8] Further, our previous theoretical calculations have shown that the delocalized MOs exist in water rings.[9-10] These studies help to understand H-bonds from the perspective of MOs at the atomic level.

The $(H_2O)_2$ dimer is the simplest water cluster and the spatial conformation benchmark for studying complex H-bonding systems. Its H-bonding conformation is frequently studied both experimentally and theoretically.[2-3, 11-13] However, more comprehensive studies are still needed since the fundamental mechanism of interaction between two water molecules is still not clearly understood. The covalent-like characteristics in the H-bonds between two 8-hydroxyquinoline molecules revealed in a recent experiment using atomic force microscopy were found to originate from both the covalent charge in H···N and the charge transferred from H to N and O,[7] which has stimulated much interest among experimentalists and theoreticians to further explore the intermolecular interaction mechanism of $(H_2O)_2$.

In this study, we present a study aiming to understand the H-bonding mechanism of $(H_2O)_2$ from the MO perspective, which allows illustration of the nature of molecular interaction.[5, 14-16] The combination of orbital morphology with orbital composition offers an intuitive visualization and a qualitative interpretation.

**Methods**

To achieve insightful MO analysis of the interaction systems, we adopted density functional theory (DFT) to obtain accurate geometric parameters and further visualization of Kohn-Sham MOs.[5, 17] We used the coupled cluster singles, doubles, and perturbative triples [CCSD(T)] levels of theory[18-20] to validate our DFT results. For the DFT calculations, we chose to use a PBE0 functional as it has been widely used in describing H-bonding interactions and is capable of offering a good performance for treating H-bonds.[21] As such, we have optimized the $(H_2O)_2$ at both CCSD(T) and PBE0 levels with an aug-cc-pvqz basis set[22] using Gaussian 09.[23] We further performed energy decomposition based on SAPT using the Molpro 2012 program.[24] The basis set superposition error, is calculated using the counterpoise method. Considering the good agreement between the SAPT(CCSD) and SAPT(DFT) results, which give quite similar energy components in nearly all cases,[25] we performed SAPT(DFT) calculations of $(H_2O)_2$ using δ(HF) correction.

**Results and Discussion**

The optimized lowest energy structure of $(H_2O)_2$ is displayed in Figure 1. For convenience, the geometrical details are given for the following discussion.

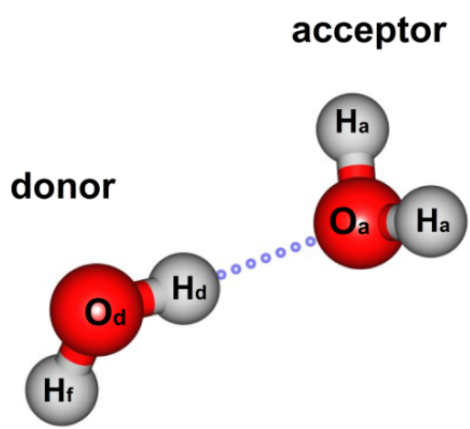

Figure 1. The equilibrium structure of $(H_2O)_2$.

The bonding property of the intermolecular interaction system is effectively revealed by MO analyses,[5, 16] as shown in Figure 2. The quantitative contributions (in percentages) from atomic orbitals to these complex MOs are also given. Here, the orbital interaction diagram of $(H_2O)_2$ is from calculations at the DFT-PBE0 level.[26-27] The PBE0 functional has been shown to give orbital diagrams consistent with results using another *ab initio* method (see Part 1 of the Supporting Information). When the contribution from a fragment orbital (FO, i.e. the MO of the water monomer) to a complex orbital is larger than 0.5%, the two energy levels respectively corresponding to FO and the complex orbital are linked in Figure 2. As is shown, two MOs (HOMO-2 and HOMO-4) clearly cross the region between the two water monomers. The HOMO-2 of $(H_2O)_2$ is formed by mixing FO HOMO-1 (82%) in the donor molecule with FO HOMO-1 (5%) and HOMO (13%) in the acceptor molecule. The HOMO-4 of $(H_2O)_2$ is formed by mixing FO HOMO-2 (95%) in the donor molecule with FO HOMO-1 (3%) and HOMO (2%) in the acceptor molecule. These two crossing MOs are mainly composed of the $2p$ orbital of O and $1s$ orbital of H of the donor molecule, with certain contribution from the acceptor molecule (see Figure 1 for notations of the atoms and Table S1 for their contributions in percentages). Additionally, we obtained a bond order of 0.03 from atom-atom overlap-weighted Natural Atomic Orbital bond order[28] analysis and 0.08 from Mayer bond order[29] analysis of the water dimer, confirming the covalent characteristics in the weak hydrogen bond in $(H_2O)_2$.

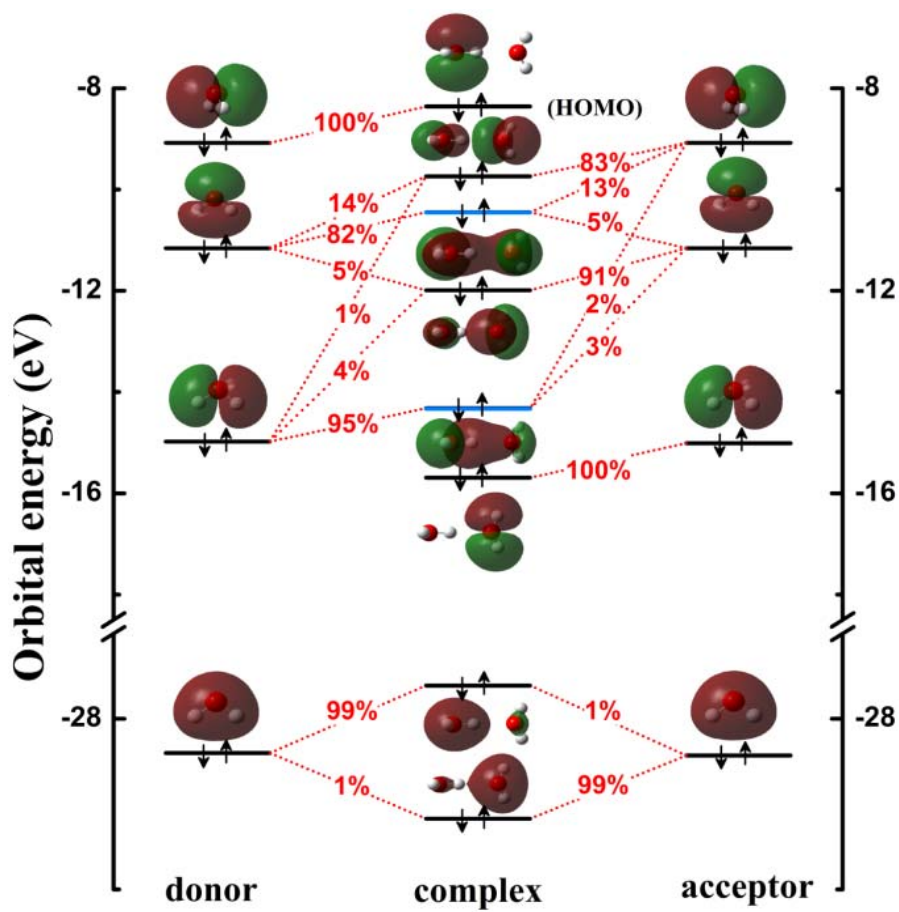

Figure 2. The orbital interaction of $(H_2O)_2$. Orbital energy levels are represented as solid bars. The bars on the left and right sides correspond to the fragment orbitals (FOs) of the two water monomers; the bars in the middle correspond to the complex orbitals of $(H_2O)_2$. The topmost solid black bars denote the highest occupied MOs (HOMOs). Blue solid bars denote two bonding MOs between the two water monomers, HOMO-2 and HOMO-4. Two corresponding bars are linked by short red dotted lines in the center of which the component percentage values (%) are given for those with the composition of a FO in a complex orbital larger than 0.5%.

To achieve further insight into the interaction between water monomers, we additionally analyzed the composition of the H-bond of $(H_2O)_2$ using the symmetry-adapted perturbation theory (SAPT) treatment. The total interaction energy ($E_{int}$) between two water molecules can be decomposed as:

$$E_{int} = E_{elec} + E_{exch} + E_{ind} + E_{disp} + \delta(HF)$$

where $E_{elec}$ describes the classical Coulomb interaction between water monomers; $E_{exch}$ is the exchange-repulsion term; $E_{ind}$ is the energy of interaction of the permanent multipole moments of one monomer and the induced multipole moments of the other. This term is interpreted as orbital interaction, and became a standard by which to determine whether a system has a covalent characteristic,[5, 30] representing the polarization of the electron density between water monomers; $E_{disp}$ is the dispersion interaction energy; the $\delta(HF)$ term is a Hartree-Fock (HF) correction for higher-order contributions to the total interaction energy obtained at HF level (further details are listed in Part 3 of the Supporting Information).

Table 1. SAPT interaction energy (kcal/mol) decomposition results for $(H_2O)_2$.

|  | SAPT(HF) | SAPT(DFT) |
|---|---|---|
| $E_{elec}$ | -8.37 | -8.10 |
| $E_{exch}$ | 7.04 | 8.04 |
| $E_{ind}$ | -1.35 | -1.37 |
| $E_{disp}$ | — | -2.41 |
| $\delta(HF)$ | -0.92 | -0.92 |
| $E_{int}$ | -3.60 | -4.76(-4.95[a]) |

[a] denotes the interaction energy calculated at CCSD(T)/aug-cc-pVQZ level.

We used the ionization potential of 0.4638 a.u.[31] in calculations with asymptotic correction. The SAPT(HF) and SAPT(DFT) interaction energy decompositions for the $(H_2O)_2$ are shown in Table 1. The consistency of the two data sets confirms the unique trend we demonstrate. It can be seen that the $E_{elec}$ term is the most important contribution to $E_{int}$, about twice as large as the total interaction energy and greater than 60% of attractive interaction energy. The $E_{disp}$ term makes certain contributions to total interaction energy. Importantly, the $E_{ind}$ term plays a non-negligible role in stabilizing the $(H_2O)_2$, which amounts to more than 10% of the attractive interaction energy and about 50% of $E_{disp}$.

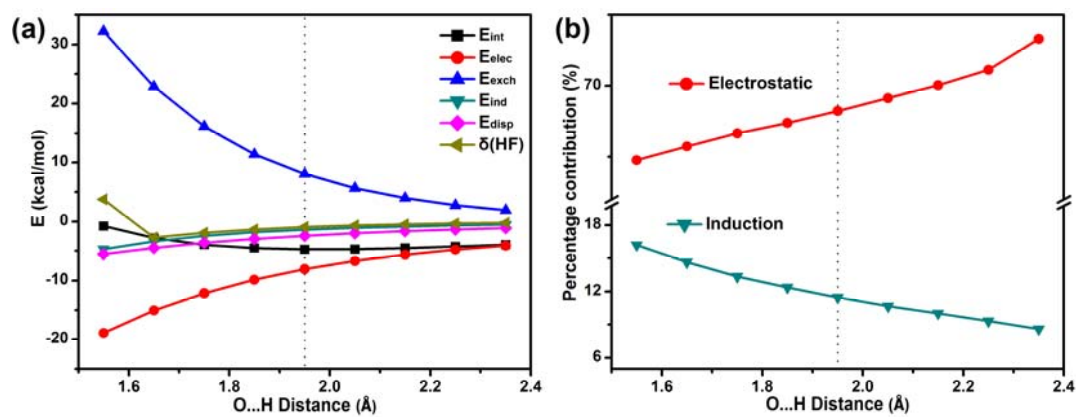

Figure 3. Energy decomposition of $(H_2O)_2$ at different O…H distances and the contribution percentages of electrostatic and induction energies to the total attractive interaction. (a) Interaction, electrostatic, exchange, induction, and dispersion energies of $(H_2O)_2$ at different O…H distances ($D_{O…H}$). (b) The percentage values represent contribution to the total attractive interactions. The black dotted lines represent the equilibrium O…H distance of $(H_2O)_2$.

The above energy decomposition analysis shows that the system has some covalent characteristics relating to the orbital overlap. $E_{ind}$ should be intermolecular distance dependent. The greater the distance between molecules, the less orbital overlap there should be, leading to a reduced $E_{ind}$. Therefore, we further analyzed the energy decomposition results at different O…H distances in order to confirm the reliability of our results. The energy decomposition result is given in Figure 3. Figure 3(a) presents the interaction, electrostatic, exchange, induction, and dispersion energies of $(H_2O)_2$ at different O…H distances ($D_{O…H}$), while Figure 3(b) shows the corresponding percentage contribution to the total attractive interactions. The results show that with the increase of the O…H distances, the contribution from the electrostatic interaction term gradually increases, and the contribution from the induction term decreases accordingly. The trend is particularly clear in Figure 3(b), where the change in the percentage contribution of the interaction energy shows that the induction decreases monotonically as the distance increases, reflecting the weakening of the covalent characteristics. This trend is reasonable and consistent with our previous report on the water tetramer.[10] The results further prove that this work may have a fundamental significance in understanding water clusters and complex H-bonding systems.

**Conclusions**

In summary, the calculated electronic structure illustrates the orbital interaction between water monomers in $(H_2O)_2$. There are two MOs crossing the $(H_2O)_2$ system along the H-bonding region. Furthermore, the energy decomposition analysis demonstrates that $E_{ind}$ is non-negligible in the interaction between two monomers. This non-negligible $E_{ind}$ suggests that the interaction has a covalent-like character,[32-33] given that orbital interaction is interpreted as induction interaction or polarization interaction.[5, 30] The results with direct bonder order calculations clearly reveal the covalent-like character of the simplest H-bonding structure of $(H_2O)_2$. Recently, the electronic structure of the H-bond has been visualized by low-temperature scanning tunneling microscope.[8] This study provides a critical insight at the atomic level for understanding H-bonding systems and the prediction of their properties.

**Acknowledgements**

We thank Dr. Bo Zhou for providing technical support. This work was supported by the National Science Foundation of China (grant number 11374004) and the Science and Technology Development Program of Jilin Province, China (20150519021JH). Z. W. also acknowledges the Fok Ying Tung Education Foundation (142001) and the High Performance Computing Center of Jilin University.

# Supplementary Information

# Covalent features in the hydrogen bond of a water dimer: molecular orbital analysis

Bo Wang[1,2], Wanrun Jiang[1,2], Xing Dai[1,2], Yang Gao[1,2], Zhigang Wang[1,2,3,*] and Rui-Qin Zhang[4,5,*]

[1]Institute of Atomic and Molecular Physics, Jilin University, Changchun 130012, China

[2]Jilin Provincial Key Laboratory of Applied Atomic and Molecular Spectroscopy (Jilin University), Changchun, 130012, China

[3]Institute of Theoretical Chemistry, Jilin University, Changchun 130023, China

[4]Department of Physics and Materials Science and Centre for Functional Photonics (CFP), City University of Hong Kong, Hong Kong SAR, China

[5]Beijing Computational Science Research Center, Beijing 100084, China.

*To whom correspondence should be addressed. E-mail: wangzg@jlu.edu.cn, Tel.: +86-431-85168817; aprqz@cityu.edu.hk, Tel.: +852-34427849.

**Contents**

**Part 1. The orbital interaction diagram of (H$_2$O)$_2$ at HF level based on CCSD(T) calculations.**

**Part 2. The contribution percentages from atomic orbitals to complex MOs.**

**Part 3. More details of interaction energy decomposition.**

**Part 1. The orbital interaction diagram of (H$_2$O)$_2$ at HF level based on CCSD(T) calculations.**

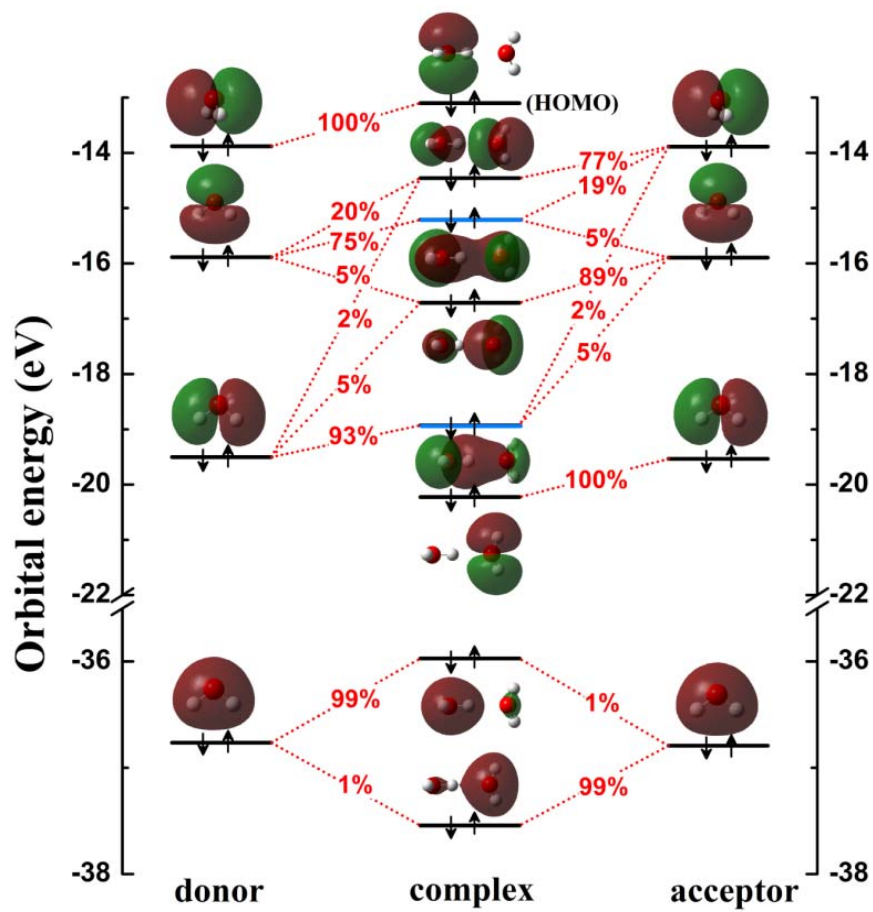

Figure S1. The orbital interaction of (H$_2$O)$_2$. Orbital energy levels are represented as solid bars. The bars on the left and right columns correspond to the fragment orbitals (FOs) of the two water monomers; the bars in the middle column corresponds to complex orbitals of (H$_2$O)$_2$. The topmost black solid bars denote the highest occupied MOs (HOMOs). Blue solid bars denote two bonding MOs between two water monomers, HOMO-2 and HOMO-4. Two corresponding bars are linked by red short dot lines and the component percentage values (%) are given for those with the composition of a FO in a complex orbital larger than 0.5%.

To clarify that the orbital diagram presented in the manuscript is consistent with results using other ab initio methods, we calculated the orbital interaction of (H$_2$O)$_2$ at HF level based on CCSD(T) calculations. When the contribution component of a fragment orbital (FO, i.e. the MO of water monomer) to a complex orbital is larger than 0.5%, the two energy levels respectively corresponding to FO and complex orbital are linked in Figure. S1. Here we found that there are two MOs (HOMO-2 and HOMO-4) clearly cross the region between the two water monomers. The HOMO-2 of (H$_2$O)$_2$ is formed by mixing FO HOMO-1 (75%) in the donor molecule with FO HOMO-1 (5%) and HOMO (19%) in the acceptor molecule. The HOMO-4 of $(H_2O)_2$ is formed by mixing FO HOMO-2 (93%) in the donor molecule and the FO HOMO-1 (5%) and HOMO (2%) in acceptor molecule.

**Part 2. The contribution percentages from atomic orbitals to complex MOs.**

Table S1. MO components of $(H_2O)_2$. The corresponding atomic labels are given in Figure 1. Results in bold denote the two crossing MOs.

| $(H_2O)_2$ (%) | $O^d$ | | $H^f$ | $H^d$ | $O^a$ | | $H^a$ | $H^a$ |
|---|---|---|---|---|---|---|---|---|
| | 2s | 2p | 1s | 1s | 2s | 2p | 1s | 1s |
| HOMO | 0.00 | 99.38 | 0.00 | 0.00 | 0.00 | 0.00 | 0.01 | 0.01 |
| HOMO-1 | 0.70 | 14.71 | 0.37 | 0.05 | 0.01 | 83.60 | 0.00 | 0.00 |
| **HOMO-2** | **8.38** | **67.21** | **3.06** | **2.56** | **0.65** | **17.39** | **0.16** | **0.16** |
| HOMO-3 | 0.91 | 6.24 | 1.09 | 0.02 | 8.89 | 76.26 | 3.08 | 3.08 |
| **HOMO-4** | **0.00** | **69.04** | **12.74** | **12.77** | **0.11** | **4.64** | **0.16** | **0.16** |
| HOMO-5 | 0.00 | 0.01 | 0.00 | 0.00 | 0.00 | 73.59 | 13.01 | 13.01 |
| HOMO-6 | 75.79 | 3.16 | 9.44 | 9.75 | 0.42 | 0.11 | 0.07 | 0.07 |
| HOMO-7 | 0.35 | 0.07 | 0.03 | 0.28 | 76.32 | 3.08 | 9.35 | 9.35 |

**Part 3. More details of interaction energy decomposition.**

In the SAPT method, the total interaction energy, $E_{int}$ is given as the sum of first-order energy ($E^1$) and second-order ($E^2$) and $\delta(\text{HF})$ term, $E^1_{pol}$ is electrostatic interaction term, $E^1_{exch}$ is exchange-repulsion term, $E^2_{ind}$ is induction term. $E^2_{ind-exch}$ is exchange-induction term, $E^2_{disp}$ is dispersion term, $E^2_{disp-exch}$ is exchange-dispersion term. The $\delta(\text{HF})$ term is a Hartree–Fock correction for higher-order contributions to the interaction energy. These interaction energy components can be calculated according to the equations (1):

$$E_{elec} = E^1_{pol}$$

$$E_{exch} = E^1_{exch}$$

$$E_{ind} = E^2_{ind} + E^2_{ind-exch} \qquad (1)$$

$$E_{disp} = E^2_{disp} + E^2_{disp-exch}$$

$$E_{int} = E_{elec} + E_{exch} + E_{ind} + E_{disp} + \delta(\text{HF})$$